\documentclass[twocolumn,trackchanges]{aastex6}
\usepackage{color}
\usepackage{graphicx}
\usepackage{hyperref}

\newcommand{\Msun}{\ensuremath{M_{\sun}}}
\newcommand{\Mstar}{\ensuremath{M_{\star}}}
\newcommand{\ha}{H\ensuremath{\alpha}}

\newcommand{\rbox}{\ensuremath{R_{\rm box}}}
\newcommand{\Lbar}{\ensuremath{L_{\rm bar}}}

\newcommand{\DeltaT}{\ensuremath{\Delta T_{\rm non-B/P}}}
\newcommand{\deltaT}{\ensuremath{\delta T_{\rm buck}}}
\newcommand{\ktwo}{\ensuremath{A_{\sigma}}}

\shorttitle{Detection of Bars in the Buckling Phase}
\shortauthors{Erwin \& Debattista}

\begin{document}

\title{Caught in the Act: Direct Detection of Galactic Bars in the Buckling Phase}

\author{Peter Erwin}
\affil{Max-Planck-Institut f\"{u}r extraterrestrische Physik,
Giessenbachstrasse, D-85748 Garching, Germany\\
Universit\"{a}ts-Sternwarte M\"{u}nchen, Scheinerstrasse 1,
D-81679 M\"{u}nchen, Germany}
\email{erwin@mpe.mpg.de}
\and
\author{Victor P. Debattista}
\affil{Jeremiah Horrocks Institute, University of Central Lancashire, Preston PR1 2HE, UK}

\begin{abstract}
The majority of massive disk galaxies, including our own, have stellar
bars with vertically thick inner regions -- so-called
``boxy/peanut-shaped'' (B/P) bulges. The most commonly suggested
mechanism for the formation of B/P bulges is a violent vertical
``buckling'' instability in the bar, something that has been seen in
$N$-body simulations for over twenty years, but never identified in real
galaxies. Here, we present the first direct observational evidence for
ongoing buckling in two nearby galaxies (NGC~3227 and NGC~4569),
including characteristic asymmetric isophotes and (in NGC~4569)
stellar-kinematic asymmetries that match buckling in simulations. This
confirms that the buckling instability takes place and produces B/P
bulges in real galaxies.  A toy model of bar evolution yields a local
fraction of buckling bars consistent with observations if the buckling
phase lasts $\sim$ 0.5--1 Gyr, in agreement with simulations.
\end{abstract}

\section{Introduction} 

Approximately 60--70\% of disk galaxies in the local universe have
stellar bars \citep[e.g.,][]{eskridge00,menendez-delmestre07}. A wide
variety of observational evidence indicates that many bars are
vertically thickened in their inner regions, appearing as ``boxy'' or
``peanut-shaped'' (B/P) bulges when seen edge-on; this includes our own
Galaxy, whose bulge is mostly if not entirely part of its bar
\citep[e.g.,][]{shen10,di-matteo14}. Edge-on galaxies with B/P bulges
show gas and stellar kinematics consistent with a rotating bar in the
disk plane
\citep{kuijken95,bureau99a,merrifield99,veilleux99,
chung-bureau04}; moderately inclined barred galaxies show isophotes
consistent with the projection of B/P bulges within the bars
\citep{bettoni94,quillen97,athanassoula06,erwin-debattista13}; and
face-on barred galaxies show kinematic and morphological signatures of
B/P bulges as well \citep{mendez-abreu08,laurikainen14}.

Recent studies suggest that failing to account for the presence of
B/P bulges can lead to significantly overestimating the luminosities and
masses of ``classical'' (spheroidal) bulges in disk galaxies
\citep{laurikainen14,athanassoula15}. This can potentially bias our
understanding of how bulges are related to other galaxy properties, including
the key correlations between supermassive black holes and bulges
\citep[e.g.,][]{kormendy13}. Understanding the formation of B/P bulges
is thus an important part of understanding and constraining models of
galaxy (and black hole) evolution.

The most frequently invoked mechanism for forming these structures is
the buckling instability of the bar, a brief but violent vertical
instability which occurs (in simulations) not long after the bar forms.
In simulations, the formation of the bar increases the radial velocity
dispersion of stars in the disk; this leads to a highly anisotropic
velocity dispersion tensor and the vertical destabilization of the bar
\citep{raha91,merritt94}. Following a phase of asymmetric vertical
buckling, the inner part of the bar settles into the more vertically
symmetric form of a B/P bulge. 

Despite over twenty years of simulations which show buckling, no
observed buckling has (yet) been reported for real galaxies, a situation
called ``puzzling'' in the review of \citet{shen16}. An alternative
model proposes that B/P bulges form by the trapping of single orbits
into vertical resonances, leading to more gradual, vertically
\textit{symmetric} growth
\citep{combes90,quillen02,debattista06,berentzen07,quillen14}. In some
simulations, the presence of gas weakens or prevents buckling, while
still allowing symmetric bar thickening
\citep{berentzen98,debattista06,berentzen07,wozniak09}. Thus, it is not
clear that real galaxies must suffer the buckling instability. There are
also no clear, strong differences due to the different formation
mechanisms in the resulting end-stage B/P bulges, making it difficult to
determine from observations of existing B/P bulges how they were formed.

In this Letter, we present evidence for ongoing buckling in the bars of
two local spiral galaxies (NGC~3227 and NGC~4569), thus demonstrating
that buckling of bars definitely occurs in real galaxies. We also argue
that the observed fraction of buckling bars at $z = 0$ is at least
broadly consistent with most (or even all) B/P bulges being the result
of buckling, if the buckling phase lasts $\sim 0.5$--1 Gyr -- as is
predicted by $N$-body simulations.


\begin{figure*}
  \centering
  
\hspace*{-0.5cm}\includegraphics[scale=1.02]{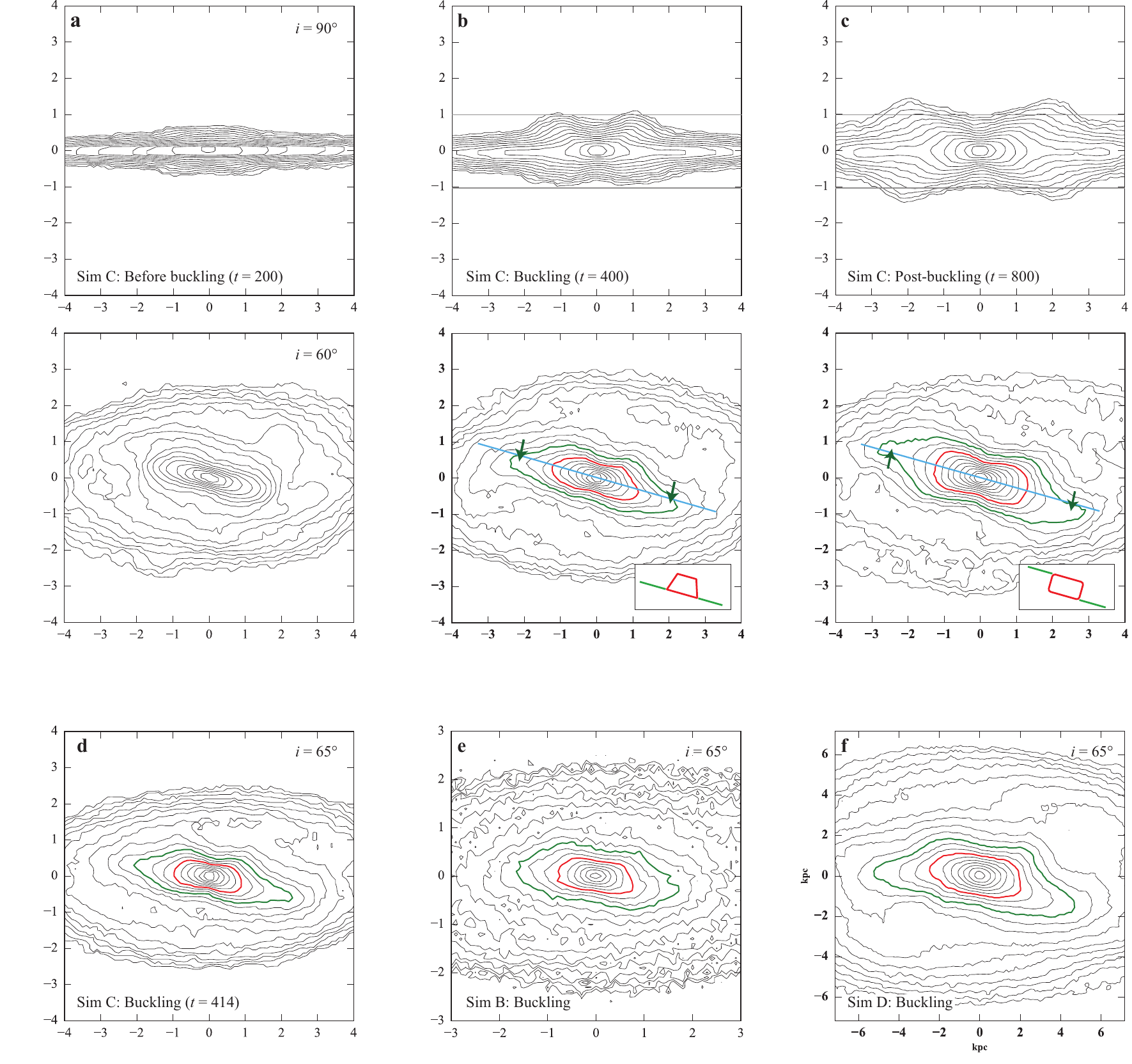}  

  \caption[]{ Edge-on and inclined views of $N$-body simulations
  before, during, and after vertical buckling. Panels a--c show
  log-scaled isodensity contours of Simulation~C for edge-on (upper
  sub-panels, with bar perpendicular to line of sight) and inclined
  views (lower sub-panels, $i = 60\arcdeg$, bar oriented 30\arcdeg{}
  from line of nodes before inclining galaxy). \textbf{a.} Before
  buckling, showing the symmetric, vertically thin bar. \textbf{b.}
  During buckling: vertical asymmetry (upper sub-panel) translates to
  asymmetric, trapezoidal inner region in lower sub-panel (red contours)
  and outer-bar ``spurs'' (green), offset in same direction (arrows)
  from major axis of inner region (cyan line) in lower sub-panel.
  \textbf{c.} After buckling: the symmetric boxy/peanut-shaped (B/P)
  bulge projects to rectangular inner contours (red) and counter-offset
  spurs (green contours, arrows); the projected bar now has 180\arcdeg{}
  rotational symmetry about the galaxy center. The small inset panels
  show cartoon versions of the basic buckling and post-buckling
  projected morphologies (red trapezoid/box + green spurs).
  \textbf{d--f.} As for lower sub-panel of \textbf{b}, but now showing
  Simulation~C later in the buckling process (\textbf{d}) and
  Simulations~B and D during their buckling phases (\textbf{e, f}); all three are
  seen with $i = 65\arcdeg$. }\label{fig:sims-pre-during-and-post}

\end{figure*}

\begin{figure*}
\centering
  \hspace*{-1.2cm}\includegraphics[scale=1.0]{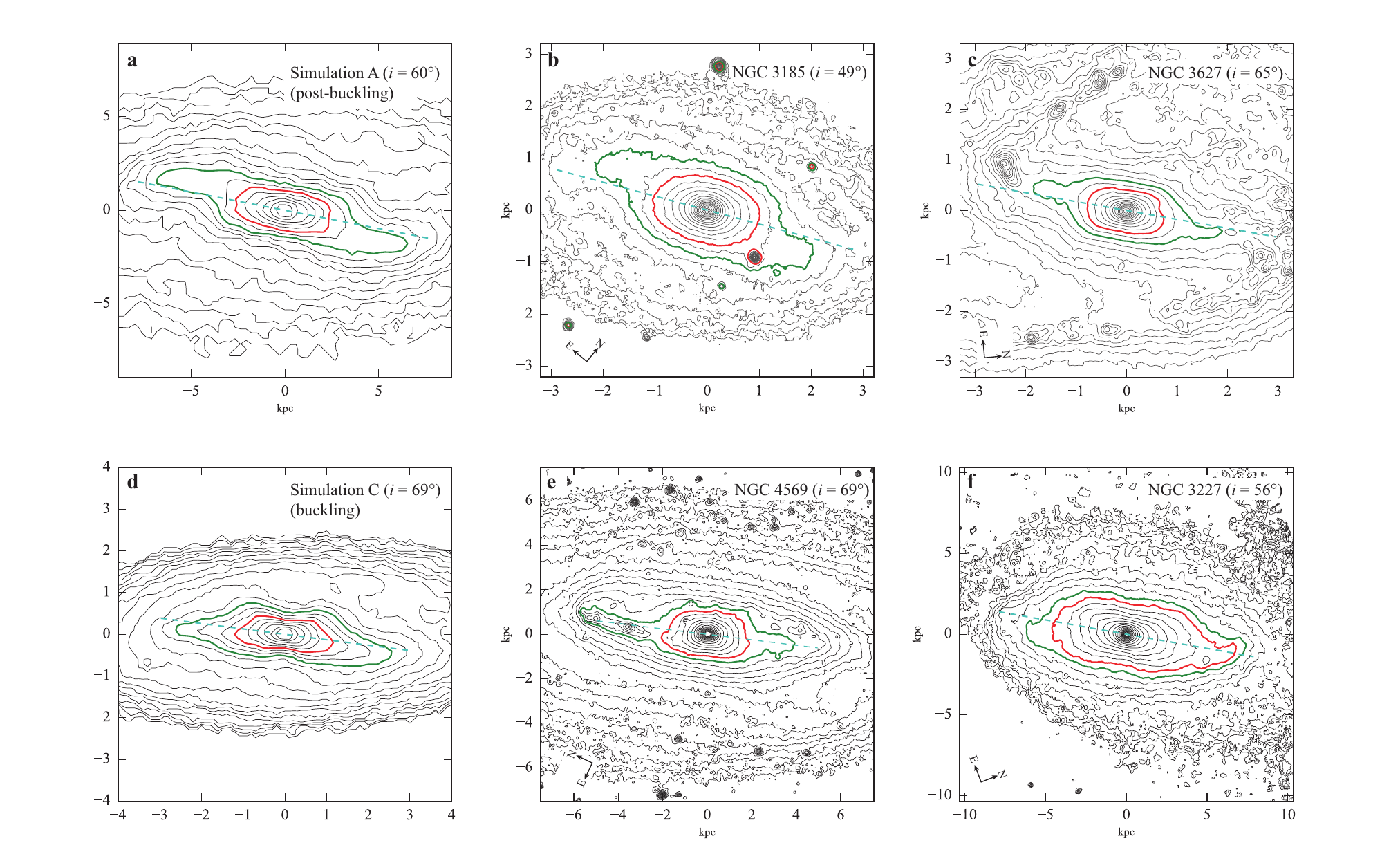} 
  
  \caption[]{$N$-body simulations during and after buckling, along with
  real galaxies seen at similar orientations. \textbf{a.} Simulation A
  after buckling: the isodensity contours show the characteristic box +
  offset spurs with 180\arcdeg{} rotational symmetry. \textbf{b.}
  $H$-band isophotes of symmetric-B/P galaxy NGC~3185
  \citep{erwin-debattista13}. \textbf{c.} \textit{Spitzer} 3.6~\micron{}
  isophotes of symmetric-B/P galaxy NGC~3627 \citep{kennicutt03}.
  \textbf{d.} Simulation C \textit{during} buckling. \textbf{e.}
  \textit{Spitzer} 3.6~\micron{} isophotes of NGC~4569 \citep{kennicutt03}.
  \textbf{f.} $K$-band isophotes of NGC~3227 \citep{mrk97}. All images
  are rotated to place disk line of nodes horizontal; all isophotes are
  logarithmically scaled.  Red contours outline approximate trapezoid
  (buckling) or boxy (post-buckling) B/P regions, green contours outline
  outer (vertically thin) bar spurs, and dashed cyan lines show
  trapezoid/box major axes. \label{fig:n3227-n4569-vs-sim}}
    
\end{figure*}

\section{Buckling and Morphology}

\subsection{$N$-body Simulations}



$N$-body simulations have long predicted that bars should buckle some time after
they form. In this Letter, we use four such simulations, three of
which were previously analyzed in \citet{erwin-debattista13}.  For
consistency, we use a similar naming scheme: ``Simulation A--C'' in this
paper correspond to ``runs A--C'' in Erwin \& Debattista. These three
simulations are described in more detail in Erwin \& Debattista and (for
simulations B and C) in \cite{sellwood09}; they included
300,000--500,000 stellar particles in the disk and softening lengths of
60~pc (Simulation A) or 0.05 natural units (B and C). Simulation D is
almost identical to model T2 of \citet{debattista16}. It is comprised of
thin+thick disks inside a dark matter halo, with 3 million stellar
particles and a softening length of 50~pc.  The two disks have roughly
the same radial velocity dispersion profile, but the thin disk has half
the height of the thick disk (versus a height ratio of one-fourth
for model T2); see Debattista et al.\ for further details.

\subsection{Projected Morphology}

\citet{erwin-debattista13} compared $N$-body simulations and real barred
galaxies to show that it was possible to identify B/P bulges in bars
even at inclinations as low as $\sim 40\arcdeg$. (The advantage of
intermediate inclinations -- e.g., $i \lesssim 70\arcdeg$ -- over more
edge-on orientations is that the bar as a whole is readily identifiable
and measurable, regardless of whether its interior is vertically thick
or not.) The key isophotal signatures are a broad, often boxy inner zone
(corresponding to the vertically thick, B/P part of the bar) and thinner
``spurs'' extending beyond (corresponding to the vertically thin outer
parts of the bar); the spurs are usually offset from the major axis of
the inner zone, so that the projected bar has 180\arcdeg{} rotational
symmetry on the sky (Fig.~\ref{fig:sims-pre-during-and-post}c,
\ref{fig:n3227-n4569-vs-sim}a).

When simulations \textit{in the buckling phase} are viewed at the same
intermediate inclinations, however, the projected bar shows
\textit{trapezoidal} inner isophotes (corresponding to the main
buckling region) and outer spurs which are offset in the \textit{same}
direction relative to the bar's observed major axis, forming a
continuation of the trapezoid's long side
(Fig.~\ref{fig:sims-pre-during-and-post}b, d--f,
\ref{fig:n3227-n4569-vs-sim}d). This morphology clearly differs from the
symmetric-box plus counter-offset spurs seen in post-buckling bars with
their symmetric B/P bulges.

\section{Detection of Buckling Bars in NGC~3227 and NGC~4569}

\subsection{Morphology}

\begin{figure*}
\centering
  \hspace*{-1.0cm}\includegraphics[scale=1.0]{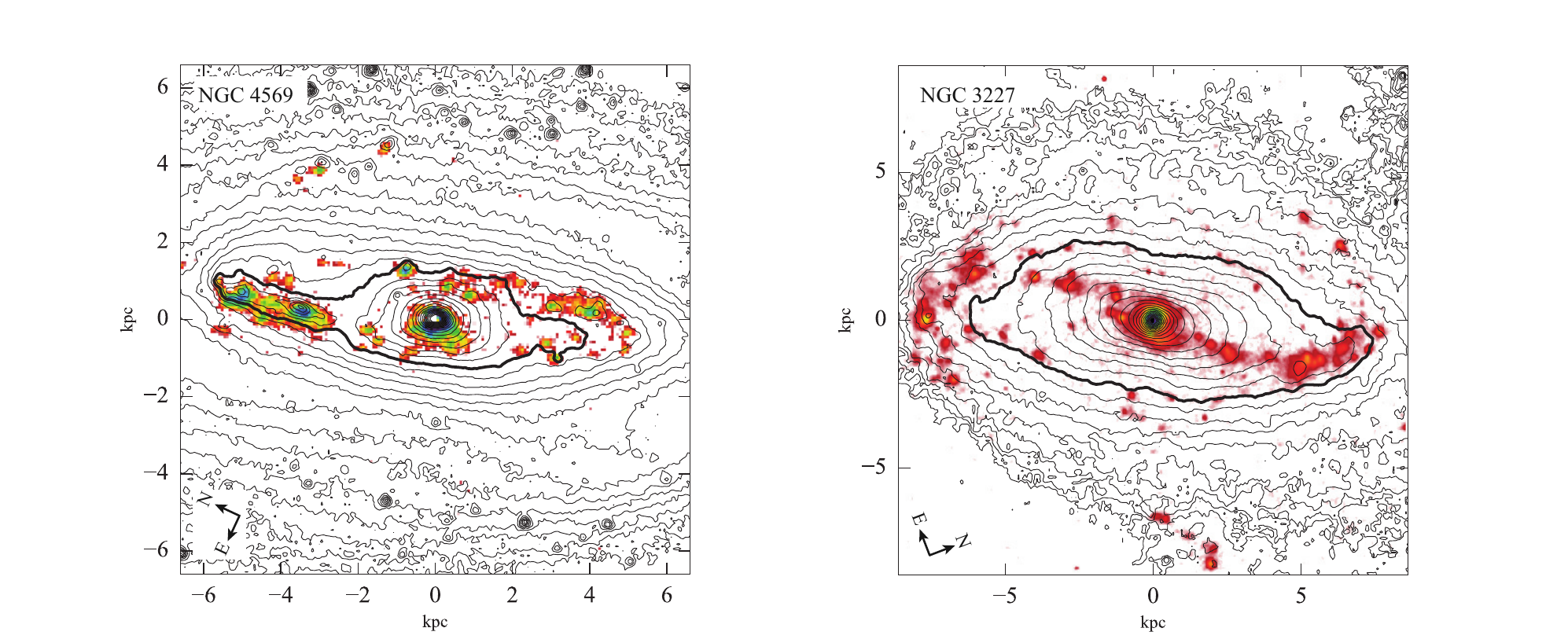}

  \caption[]{Star-formation and near-IR morphology for galaxies with
  buckling bars. We show logarithmically scaled near-IR isophotes (black
  contours; left: \textit{Spitzer} 3.6~\micron{} from \citealt{kennicutt03};
  right: $K$-band from \citealt{mrk97}) and \ha{} emission (color; left:
  \citealt{knapen04}; right: \citealt{koopmann01}); thicker contours
  outline spurs in each bar. Galaxies are rotated to place disk major
  axes horizontal. Star formation enhances the northern spurs in each
  galaxy, but is unrelated to the southern spurs. All four spurs are
  thus primarily due to the underlying stellar structure of the bars,
  not recent star formation. \label{fig:ed-sf}}

\end{figure*}

As part of a survey of local barred galaxies with favorable orientations
for detecting and measuring projected B/P bulges (Erwin \& Debattista,
in prep), we have found two examples of bars with morphologies
indicating that they are currently in the buckling phase: NGC~3227 and
NGC~4569 (M90). In both galaxies, the inner or middle region of
the bar shows quasi-trapezoidal isophotes (most clearly in
NGC~4569), while the outer-bar spurs are offset in the same direction
and connect to the long side of the trapezoid
(Fig.~\ref{fig:n3227-n4569-vs-sim}e,f). This morphology is a good match
to the general appearance of projected $N$-body bars in the buckling
phase. Although both galaxies are currently experiencing star formation
within their bars, comparison of their near-infrared morphologies with
published \ha{} images allows us to rule out the possibility that recent
star formation is responsible for the overall morphology
(Fig.~\ref{fig:ed-sf}). 

The best case is probably NGC~4569, which shows a symmetric
inner trapezoid and clear offset spurs; it differs from the simulations
(Figure~\ref{fig:sims-pre-during-and-post}) in having a relatively
compact buckling region, with a half-length $\rbox \sim$ 1.9 kpc
(measured along its long axis), quite small relative to the bar as a
whole ($\rbox/\Lbar \approx 0.29$, where $\Lbar =$ half-length of the
bar). The trapezoidal region in NGC~3227 is much larger ($\rbox = 4.7$
kpc, $\rbox/\Lbar = 0.54$), with some resemblance to Simulation~B
(Figure~\ref{fig:sims-pre-during-and-post}e); the interior of
the trapezoid appears less symmetric, which may be partly due to
emission from recent star formation (Figure~\ref{fig:ed-sf}).
Nonetheless, the outer isophotes show the characteristic offset-spurs
pattern.  The relative sizes of the buckling regions in these two
galaxies fall near the lower and upper limits for the range of
\textit{symmetric} B/P bulge sizes seen in local barred galaxies:
$\rbox/\Lbar = 0.26$--0.58, with a mean of 0.38
\citep{erwin-debattista13}.

\subsection{Stellar Kinematics}

The buckling phase in our simulations is also accompanied by asymmetries
in the stellar velocity dispersion $\sigma$, measured along the major
axis of the bar.  For simulated bars before and after the buckling
phase, the dispersion is roughly symmetric about the center: $\sigma(-x)
/ \sigma(+x) \approx 1$, where $x$ is measured along the bar major axis.
During the buckling phase, however, the dispersion becomes strongly
\textit{asymmetric}. To quantify this, we computed a mean asymmetry 
measure:
\begin{equation}
\ktwo(R) = |1 - \frac{1}{N} \sum_{i=1}^{N} \sigma(-x_{i}) / \sigma(+x_{i}) | ,
\end{equation}
where the dispersion is measured in $N$ radial bins $x_{i}$ on either
side of the galaxy center, out to a maximum
distance $R = |x_{N}|$. Symmetric dispersion profiles have values of
\ktwo{} close to 0, while more asymmetric profiles have larger values.

To determine \ktwo{} for a simulation, we first matched the
orientation of NGC~4569 (the only one of the two galaxies with sufficiently
extended stellar kinematics) by rotating the simulation so that the bar was
$\Delta$PA = 26\arcdeg{} away from the line of nodes, and then inclining
it by $69\arcdeg$ about the line of nodes. We measured velocity
dispersions as the square root of the variance of particle line-of-sight
velocities, using a slit width of 0.2 and evenly spaced bins of radial
size 0.05 (simulation units) along the major axis of the projected
simulation. For observational comparison, we used published velocity
dispersion data for NGC~4569 and ten barred galaxies with known (symmetric)
B/P bulges which had inclinations and bar orientations similar to the two
buckling-bar galaxies; these galaxies and the sources of kinematic data
are listed in Table~\ref{tab:kinematics}. For all but two of the
symmetric-B/P galaxies, we use long-slit data, if the position
angle of the slit was within 30\arcdeg{} of the bar major axis. For
NGC~3627 and NGC~4293, we extracted pseudo-long-slit measurements along
the bar major axes from published integral-field-unit (IFU)
velocity-dispersion data \citep{falcon-barroso06,dumas07}. Finally, we
derived velocity-dispersion measurements for NGC~4569 from the
IFU data of \citet{cortes15}.

\begin{figure*}
\centering

  \hspace*{-0.08cm}\includegraphics[scale=1.055]{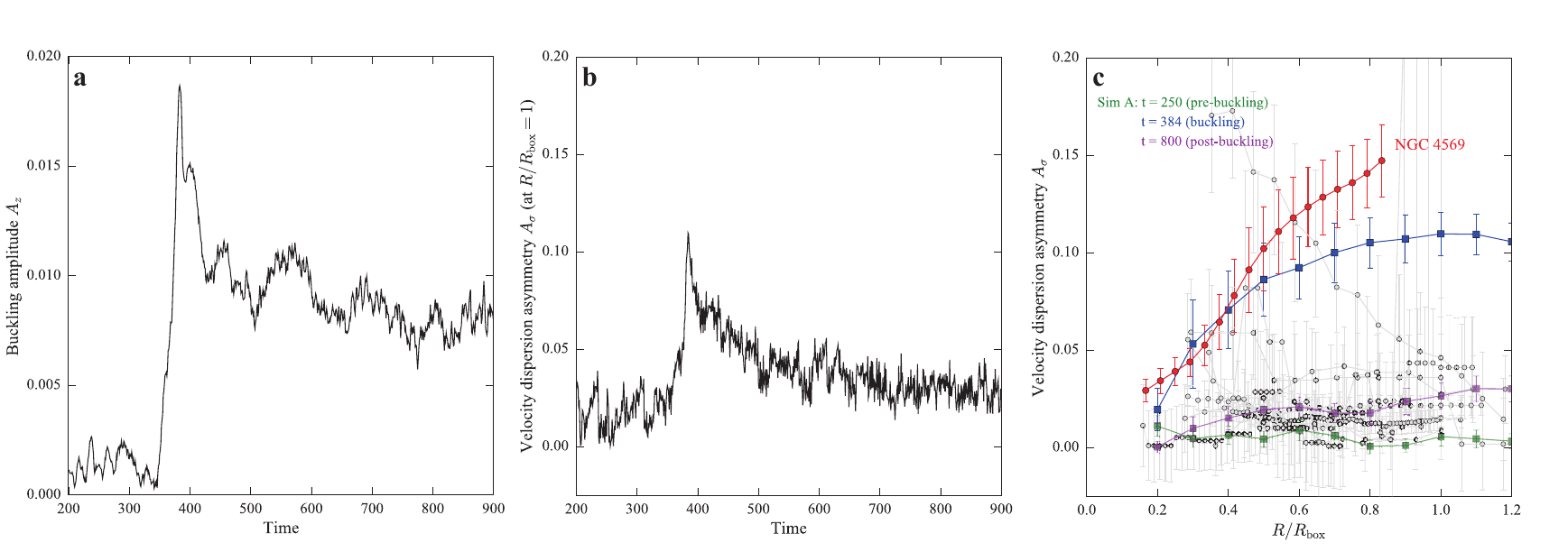} 
  
  \caption[]{Stellar velocity dispersion asymmetry along the bar in
  simulated and real galaxies. \textbf{a.} Evolution of vertical $m = 2$
  buckling amplitude $A_{z}$ \citep{debattista06} in $N$-body simulation
  C. Buckling begins at $t \approx 350$, peaks at $t \approx 385$, and
  has largely ceased for $t > 500$. \textbf{b.} Evolution of
  velocity-dispersion asymmetry $A_{\sigma}$ for the simulation (viewed
  with same orientation as NGC~4569), integrated along the bar major
  axis out to $R/\rbox = 1$. \textbf{c.} Radial trends in $A_{\sigma}$,
  averaged out to increasing radii $R$ along the bar for simulation C,
  galaxies with symmetric B/P bulges, and NGC~4569. Symmetric B/P bulges
  (gray) show $A_{\sigma}$ close to 0, especially for larger values of
  $R/\rbox$, agreeing with post-buckling state of the simulation
  (magenta). But in NGC~4569 (red), $A_{\sigma}$ increases with radius,
  similar to the simulation \textit{during} buckling (blue). Error bars
  are standard deviations from 1000 rounds of bootstrap
  resampling.\label{fig:kinematics}}
  
\end{figure*}

Fig.~\ref{fig:kinematics}b shows the time evolution of $\ktwo$ in
simulation C, with $R$ equal to the
size of the boxy/trapezoidal region once it has formed ($R = \rbox
\approx 1$ in simulation units). The onset of buckling at $t \approx
350$ (Fig.~\ref{fig:kinematics}a) is accompanied by rapid growth in
the dispersion asymmetry (Fig.~\ref{fig:kinematics}b), which gradually
returns to near symmetry ($\ktwo < 0.05$) as the inner part of the
post-buckled bar settles into a mature B/P bulge.
Fig.~\ref{fig:kinematics}c shows \textit{radial} trends in $\ktwo(R)$ for NGC~4569
and the ten symmetric-B/P galaxies, and also for representative times
from simulation C. Galaxies with symmetric B/P bulges (gray symbols)
show minimal asymmetry (in almost all cases $\ktwo < 0.05$), especially
for larger $R$, as do the pre- and post-buckling simulation profiles.
For NGC~4569 (red symbols), however, the asymmetry \textit{increases} as
$R$ approaches the size of the buckling region -- just as in the
simulation during the buckling phase (blue symbols).

\section{The Evolution of Barred Galaxies and the Frequency of Buckling}\label{sec:model}

We are currently studying a sample of local barred galaxies in order to
identify and measure B/P bulges (Erwin \& Debattista, in prep). This
sample is diameter- and distance-limited ($D_{25} \leq 3.0\arcmin$ and
$D \lesssim 30$ Mpc, to ensure adequate resolution of bar interiors) and
includes a total of 84 barred S0--Sd galaxies which have inclinations
and orientations that maximize the detectability of B/P bulges ($i =
40$--70\arcdeg, deprojected $\Delta$PA between bar and disk major axis $<
60\arcdeg$). All galaxies were examined (using near-IR images) for the
morphology of B/P bulges \citep{erwin-debattista13}. We find B/P bulges
to be extremely common in high-stellar-mass galaxies: they are present
in $\sim 80$\% of the 44 barred disks with $\log \Mstar \gtrsim 10.4$
\Msun.

The two buckling-bar galaxies were identified as part of this sample,
and are in the high-stellar-mass ($\log \Mstar \ge 10.4$) subsample. The
frequency of observed buckling in local, high-mass barred galaxies is
thus $f_{\rm buck} = 4.5^{+4.3}_{-2.3}$\%. Is this frequency
high or low? Put another way, is it consistent with the possibility that
many -- or even \textit{all} -- B/P bulges are the result of the
buckling instability?

To test this hypothesis, we adopted a toy galaxy-evolution model in
which the fraction of disk galaxies with visible bars is a linear
function of redshift, equal to $F_{0}$ at $z = 0$ and decreasing to 0 at
redshift $z_{i}$; we based this on the observed evolution of bar
fraction with redshift in spiral galaxies
\citep[e.g.,][]{sheth08,cameron10,melvin14} and recent
cosmologically-motivated simulations of bar formation and growth
\citep{kraljic12}. (We note that the simulated galaxies in Kraljic et
al.\ all have $\log \Mstar > 10.2$, consistent with the local high-mass
subsample we are considering, and that the high-redshift observational
studies have similar lower limits on \Mstar.) Following the pattern
observed in $N$-body simulations, we assumed that there is a delay
between bar formation and buckling, equal to $\DeltaT$ Gyr, followed by a
visible buckling phase lasting $\deltaT$ Gyr, and ending with a
permanently B/P-hosting bar. 

We performed a Markov Chain Monte Carlo analysis of this model,
computing the likelihood as the product of individual binomial
likelihoods for the $z = 0$ fraction of galaxies with bars $f_{\rm
bar}$, the fraction of bars with B/P bulges $f_{\rm B/P}$, and the
fraction of bars currently buckling $f_{\rm buck}$, compared with the
observed counts in our local high-mass subsample (Erwin \& Debattista,
in prep.); we assumed a flat prior for $0 < z_{i} < 2$ and flat priors
for values $\ge 0$ for the other parameters. We used the \texttt{emcee}
ensemble sampler code of \citet{foreman-mackey13}, with 50 separate
chains and 500 steps per chain, discarding the first 150 steps in each
chain. After marginalizing over $F_{0}$ and $z_{i}$ as nuisance
parameters, we find $\DeltaT = 2.2^{+1.3}_{-1.1}$ Gyr and
$\deltaT = 0.8^{+0.7}_{-0.4}$ Gyr (medians and 68\% confidence
intervals). These are in good agreement with values from $N$-body
simulations, which typically find $\DeltaT \sim 1$--2 Gyr and $\deltaT
\sim 0.5$--1 Gyr
\citep{martinez-valpuesta04,martinez-valpuesta06,saha13}. 

We conclude that the observed frequency of buckling bars at $z = 0$ is
consistent with the predictions of $N$-body simulations, and that the
buckling instability could plausibly account for most if not all
instances of B/P bulges in massive disk galaxies in the local universe.
(We note, however, that given the small numbers involved, our
observations do not rule out the possibility that some of B/P bulges
could be the result of alternate, symmetric growth mechanisms.) Direct
confirmation of this would involve imaging of bars in the buckling phase
at higher redshifts. Our simple model predicts a maximum buckling
fraction of $f_{\rm buck} \sim 0.4$ at $z \sim 0.7$; near-IR detection
of buckling in large bars ($\Lbar \gtrsim 4$ kpc) at this redshift is
feasible with the upcoming \textit{James Webb Space Telescope}.

\acknowledgements

We are pleased to thank Guillermo Blanc, Juan Cort{\'e}s, Eric Emsellem,
and J{\'e}sus Falc{\'o}n-Barroso for generously making their kinematic
data available; we also thank the referee (Juntai Shen), Jerry Sellwood,
Roberto Saglia, Jairo M{\'e}ndez-Abreu, and David Trinkle for useful
comments on earlier drafts. V.P.D.\ was supported by STFC Consolidated
grant \#~ST/M000877/1. V.P.D.\ also acknowledges with great pleasure the
support of the Pauli Center for Theoretical Studies, which is supported
by the Swiss National Science Foundation (SNF), the University of
Z{\"u}rich, and ETH Z{\"u}rich.

This research has made use of the NASA/IPAC Extragalactic Database
(NED), which is operated by the Jet Propulsion Laboratory, California
Institute of Technology, under contract with the National Aeronautics
and Space Administration.

\facility{Spitzer (IRAC)}


\begin{deluxetable}{lrrrrrrr}
\tablewidth{0pt}
\tablecaption{Stellar Kinematic Data\label{tab:kinematics}} 
\tablecolumns{8}
\tablehead{
\colhead{Galaxy} & \colhead{$i$} & \colhead{PA} & \colhead{Bar PA} &
\colhead{$\Delta$PA} & \colhead{Slit PA} & \colhead{\rbox} &
\colhead{Source} \\
 & $[\arcdeg]$ & $[\arcdeg]$ & $[\arcdeg]$ & $[\arcdeg]$ & $[\arcdeg]$ & $[\arcsec]$ & }

\startdata
NGC~615   & 69 & 158 & 162 &  11 & 155  & 11  & 1\\
NGC~1023  & 66 &  69 &  58 &  26 &  87  & 35  & 2\\
NGC~2962  & 53 &   7 & 168 &  30 & 175  & 17  & 3\\
NGC~3031  & 58 & 150 & 160 &  18 & 137  & 97  & 2\\
NGC~3627  & 65 & 175 & 161 &  31 & 161  & 18  & 4\\
NGC~4293  & 63 &  65 &  77 &  25 &  75  & 17  & 5\\
NGC~4429  & 63 &  90 & 106 &  32 &  93  & 45  & 6\\
NGC~4569  & 69 &  25 &  15 &  26 &  15  & 24  & 7\\
NGC~4725  & 42 &  40 &  50 &  13 &  35  & 63  & 8\\
NGC~6744  & 52 &  21 & 177 &  36 &   0  & 29  & 9\\
NGC~7531  & 59 &  22 &   7 &  27 &  15  & 14  & 1\\
\enddata 

\tablecomments{For each galaxy we list its name, inclination, position
angle (on the sky) of the disk, position angle of the bar, deprojected
angle between the bar and disk major axis, position angle of the slit
(or pseudo-slit for IFU data), radius of the B/P bulge \rbox{} within the bar,
and sources of kinematic data used for Figure~\ref{fig:kinematics}.
References for kinematic data: 1 = \citet{pizzella04}; 2 =
\citet{fabricius12}; 3 = \citet{simien00}; 4 = \citet{dumas07}; 5 =
\citet{falcon-barroso06}; 6 = \citet{simien97b}; 7 = \cite{cortes15}; 8
= \citet{heraudeau99}; 9 = \citet{bettoni97}.}

\end{deluxetable}

\end{document}